\documentclass[aps,prd,superscriptaddress,twocolumn,nofootinbib]{revtex4-2}

\usepackage{amsfonts,amssymb,amsmath} 
\usepackage{mathrsfs} 
\usepackage{color} 
\usepackage{graphicx}
\usepackage{dcolumn}
\usepackage{bm}

\newcommand{\B}[1]{{\mathbb #1}} 
\newcommand{\C}[1]{{\mathcal #1}}

\newcommand{\BF}[1]{{\mathbf #1}} 

\newcommand{\Tr}{\mathop{\rm Tr}} 
 
\newcommand{\half}{\frac 12} 
 
\newcommand{\quarter}{\frac 14}

\newcommand{\Slash}[1]{{\ooalign{\hfil#1\hfil\crcr\raise.167ex\hbox{/}}}}

\begin{document}

\title{Standard Model Higgs inflation supplemented by minimal dark matter}


\author{Shinsuke Kawai}
\email[]{kawai@skku.edu}
\affiliation{Department of Physics, 
Sungkyunkwan University,
Suwon 16419, Republic of Korea}
\author{Nobuchika Okada}
\email[]{okadan@ua.edu}
\affiliation{Department of Physics and Astronomy, 
University of Alabama, 
Tuscaloosa, AL35487, USA}
\author{Qaisar Shafi}
\email[]{qshafi@udel.edu}
\affiliation{Bartol Research Institute, Department of Physics and Astronomy,
University of Delaware, Newark, DE 19716, USA}

\date{\today}

\begin{abstract}

Renormalisation group analysis with the present measurements of the top quark mass $m_t = 172.69\pm 0.30$ GeV \cite{ParticleDataGroup:2022pth} indicates that the Standard Model (SM) Higgs potential becomes unstable at energy scales $\sim 10^{10}$ GeV. This may be interpreted as hinting at new particles at high energy. The minimal extension of the SM that can avoid this instability while leaving the SM Higgs as the sole scalar particle of the theory is obtained by adding suitable fermions to the SM. These fermions are good dark matter candidates and the model is known as the minimal dark matter model.
We revisit the inflationary scenario based on the minimal dark matter model, taking into account updated parameter constraints and recent understanding of reheating dynamics. We explore the model with different values of the right-handed neutrino mass and find that the cosmological prediction is insensitive to such details. We obtained a spectral index of the cosmic microwave background $n_s=0.9672$ and a tensor-to-scalar ratio $r=0.0031$ as a robust prediction of this scenario.

\end{abstract}

    
\maketitle
 
\section{Introduction}

The absence of experimental signals beyond the Standard Model (SM) in the Large Hadron Collider may indicate a vast desert of hierarchy above the electroweak scale. 
In particular, the SM Higgs may be the only scalar particle in Nature, possibly up to the Planck scale.
In cosmology, inflation is essential as it solves conceptual issues such as the flatness problem and the horizon problem, and also the cosmic microwave background (CMB) and the large scale structure of the Universe are believed to be seeded by quantum fluctuations generated during inflation, which is driven by a scalar field (inflaton).
It is thus natural to suppose that the SM Higgs boson may actually be identified as the inflaton, a possibility which has been a focus of much attention and actively investigated for several years \cite{CervantesCota:1995tz,Bezrukov:2007ep}.
A strength of the Higgs inflation model, besides its simplicity, is that the spectrum of CMB predicted by this model turns out to be in very good agreement with recent observations by the Planck satellite \cite{Planck:2018jri}.
The model has some weaknesses though\footnote{
This model necessarily involves a large dimensionless parameter $\xi\sim 10^4$ as the nonminimal coupling between the Higgs field and the scalar curvature, which has led some to fret over the issue of unitarity \cite{Barbon:2009ya,Burgess:2009ea,Burgess:2010zq,Lerner:2009na,Lerner:2010mq,Hertzberg:2010dc}.
Here we accept the view \cite{Barvinsky:2009ii,Bezrukov:2010jz} that the cut off scale is field dependent and hence violation of unitarity can be avoided during inflation (see \cite{Barbon:2015fla} for a counter argument).
}.
Naturally, some are inherited from the problems of the SM itself: there is no obvious dark matter candidate in the scenario of Higgs inflation, and a mechanism to generate the small nonvanishing neutrino masses needs to be implemented separately. 
Another notable issue concerns the instability of the Higgs potential \cite{Froggatt:1995rt}.
With the central values
$m_t = 172.69$ GeV and $m_h = 125.25$ GeV for the top quark mass and the Higgs boson mass \cite{ParticleDataGroup:2022pth}, the Higgs potential is known to exhibit instability above energy scales $\sim 10^{10}$ GeV, and thus cannot be used to describe the dynamics behind inflation that is expected to take place at higher energy scales.
In typical scenarios of Higgs inflation, a much smaller value of $m_t$ which is over 3-$\sigma$ from the central value\footnote{
The evaluation of the top quark pole mass depends significantly on the approach of the analysis.
For example, cross-section based indirect measurements give $170.5\pm 0.8$ GeV \cite{CMS:2019esx}, for which the Higgs potential is barely stable. 
Even in this case the prediction of our model presented below stays robust. 
} is assumed to avoid the instability.
Future experiments, such as the International Linear Collider \cite{Bambade:2019fyw}, should confirm whether the instability exists or not, but for the moment let us accept the current central value, in which case the instability indicates that some part of the assumptions needs to be modified, if the SM Higgs field is to serve as the inflaton.

One of the simplest solutions to overcome this instability problem is to incorporate some extra fields beyond the SM and modify the renormalisation group flow of the Higgs quartic coupling.
It is also preferable to assume that these extra fields provide the dark matter of the Universe, in which case they must be weakly interacting, massive, either bosonic or fermionic particles.
Such a {\em minimal dark matter} model \cite{Cirelli:2005uq} has been investigated, and constraints on and detectability of the candidate fields have been studied.

In \cite{Okada:2015zfa}, two of the present authors proposed an inflationary model based on the minimal dark matter scenario, with fermionic dark matter and right-handed neutrinos.
The model contains only one scalar field, the SM Higgs, that serves as the inflaton and is free from the various shortcomings of the original Higgs inflation model.
Below we investigate this model further.
Based on our recent understanding of the reheating mechanism, we reanalyse the reheating process of this model and show that the parametric uncertainly concerning this process is entirely removed.
We also point out that one of the scenarios proposed in \cite{Okada:2015zfa}, the $SU(2)_L$ triplet fermion dark matter scenario, is not viable any longer as the recent dark matter search experiments disfavour that case.
The other, quintet fermion dark matter scenario is found to be viable, and this inflationary scenario turns out to be very predictive and can be tested by near future experiments.

In the next section we review the minimal scenario with fermionic dark matter. 
We will not discuss the bosonic dark matter case since we wish to adhere to the original assumption that the SM Higgs is the only candidate inflaton field.
We analyse the inflationary model based on this scenario in Sec.~\ref{sec:inflation}, using the renormalisation group improved effective potential of the Higgs field and imposing the consistency condition for the reheating process.
We conclude with comments in Sec.~\ref{sec:final}.
The renormalisation group equations that we solved are summarised in the Appendix.

\section{Minimal fermionic dark matter and right-handed neutrinos \label{sec:model}}

The idea behind the minimal dark matter scenario is to supplement the SM with an $SU(2)_L$ multiplet which is electrically neutral and provides the relic dark matter abundance \cite{Cirelli:2005uq}.
Through the $SU(2)_L$ gauge coupling, the new multiplet contributes positively to the beta function of the Higgs self coupling, thereby rescuing the Higgs potential from the instability.
The multiplet is assumed to have an odd parity under a ${\B Z}_2$ symmetry which guarantees its stability.
Various candidates of $SU(2)_L$ multiplets are examined in \cite{Cirelli:2005uq}, and as previously mentioned, we are interested in fermionic ones as we wish to have a model in which the SM Higgs doublet is the only scalar field.
Gauge invariance and the ${\B Z}_2$ parity enforce the dark matter fermion to interact only through the electroweak interaction, and once its mass is given, the thermal production rate is fixed and the relic abundance can be computed.
We assume a zero hypercharge for the multiplet since severe constraints on nonzero hypercharge exist from the direct dark matter search experiments \cite{XENON:2018voc,LZ:2022ufs}. 

Among the $SU(2)_L$ fermion multiplets considered in \cite{Cirelli:2005uq}, the simplest ones are the triplet and the quintet\footnote{
A very large representation is not suitable for inflation since the gauge coupling constant tends to blow up quickly.
}.
Currently the strongest constraints are obtained from the indirect dark matter detection experiments. 
The $SU(2)_L$ fermion multiplets are severely constrained by the AMS-02 cosmic ray antiproton data \cite{AMS:2016oqu}, and it is concluded in \cite{Cuoco:2017iax} that the triplet fermion of the minimal dark matter scenario is strongly disfavoured.
The quintet fermion is still compatible with the AMS-02 data, and we thus only consider this case below.
We choose the mass of the quintet fermion so that its relic abundance coincides with the dark matter abundance 
$\Omega_{\rm DM}h^2 = 0.120$ determined by the Planck satellite measurements \cite{Planck:2018vyg}.
We adopt $M_{\rm DM} = 13.6$ TeV from a recent computation \cite{Bottaro:2021snn}.

\begin{table}[t]
\begin{tabular}{c|cccc}
  &~~~$SU(3)_c$&~~~$SU(2)_L$&~~~$U(1)_Y$~~~&~~~Mass\\
  \hline\\
  $\psi_{\rm DM}$&${\BF 1}$&${\BF 5}$&$0$& 13.6 TeV\\\\
  $N_R^i$&${\BF 1}$&${\BF 1}$&$0$&$M_{R}$\\\\
  \hline
\end{tabular}
\caption{Representations, charges and masses of the particles added to the SM. 
The index $i=1,2,3$ is the generation of the right handed neutrinos.
The mass $M_R$ is the largest component of the Majorana mass matrix that gives dominant contributions to the renormalisation group flow.
}
\label{tab:Contents}
\end{table}

We also include three families of SM singlet fermions (the right handed neutrinos) $N_{R}^i$, $i = 1, 2, 3$ so that the small nonvanishing (left handed) neutrino masses are generated via the Type I seesaw mechanism.
The list of particles added to the SM are summarised in Table~\ref{tab:Contents}.
The Lagrangian thus includes terms
\begin{align}\label{eqn:Lag}
  {\C L}\supset -y^{ij}\overline{\ell_i} N_{Rj} H
  -M_R^{ij}\overline{N_{Ri}^c} N_{Rj}
  -M_{\rm DM}\Tr \left[\overline{\psi_{\rm DM}^c}\psi_{\rm DM}\right],
\end{align}
where $y^{ij}$ is the $3\times 3$ Dirac Yukawa coupling, $\ell_i$ are the SM lepton doublets, $H$ is the SM Higgs doublet, and $\psi_{\rm DM}$ is the quintet dark matter fermion. 
Integrating out the right handed neutrinos, one obtains the seesaw formula \cite{Minkowski:1977sc,Yanagida:1979as,GellMann:1980vs,Mohapatra:1979ia} for the left handed neutrino mass $M_\nu$ below the electroweak scale,
\begin{align}
	M_\nu^{ij} = \frac{v^2}{2} y^{ik}(M_R)^{-1}_{k\ell}y^{j\ell},
\end{align}
where $v=246$ GeV is the Higgs doublet vacuum expectation value.

The right handed neutrinos contribute negatively to the beta function of the Higgs quartic coupling, as expected, and therefore deteriorate the stability of the Higgs potential.
The effect is dominated by the largest Dirac Yukawa coupling.
Thus, we consider only the dominant component of the Dirac Yukawa coupling $y_D$ and the heaviest right handed neutrino mass $M_R$.
%
%
For numerics, we employ the relation 
$S_\nu v^2/(2M_R)=10^{-10}$ GeV for 
$S_\nu \equiv y_D^\dag y_D$, 
which is obtained from the seesaw relation above with the neutrino mass scale evaluated from the atmospheric neutrino oscillation data.

\begin{figure}
\includegraphics[width=85mm]{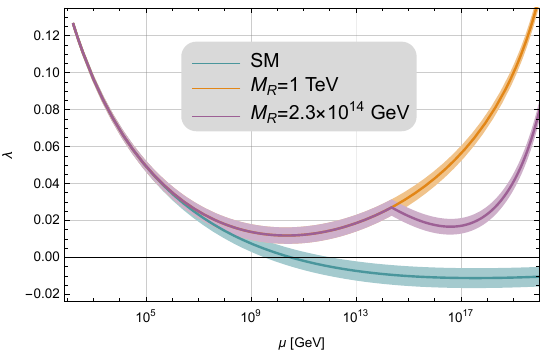}%

\includegraphics[width=83mm]{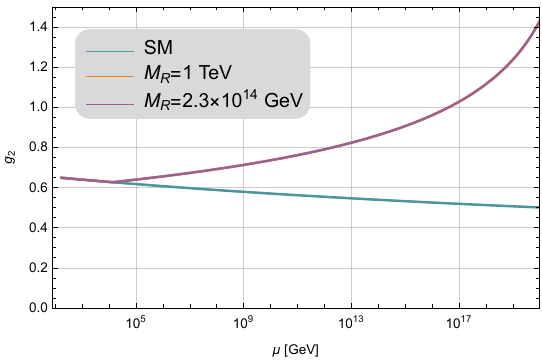}%
\caption
{\label{fig:RGflow}
Evolution of the Higgs quartic coupling $\lambda$ (the upper panel) and the $SU(2)_L$ gauge coupling $g_2$ (the lower panel).
The renormalisation group equations are listed in the Appendix.
The orange and purple curves are the minimal dark matter model with the right handed neutrino mass $M_R=1$ TeV and $2.3\times 10^{14}$ GeV respectively; the two curves overlap for $\mu<2.3\times 10^{14}$ GeV in the upper panel and for the whole range in the lower panel. 
The dark matter fermion mass is chosen to be $M_{\rm DM} = 13.6$ TeV in both cases.
The teal curve is the SM case for comparison.
The shades represent the maximum deviation arising from the 1-$\sigma$ uncertainties.
In the lower panel the uncertainties are within the line width.
}
\end{figure}

The presence of the quintet fermion modifies the renormalisation group flow above its mass, $\mu > M_{\rm DM}$, and likewise the right handed neutrinos modify the flow above $M_R$.
The full set of equations we solve and the boundary data we use are given in the Appendix.
The renormalisation group flows of the Higgs quartic coupling $\lambda$ and the $SU(2)_L$ gauge coupling $g_2$ are shown in Fig.~\ref{fig:RGflow} for the cases of the (dominant) right-handed neutrino mass $M_R=1$ TeV and $2.3\times 10^{14}$ GeV.
The SM case is also shown for comparison.
The presence of the quintet fermion renders the Higgs potential stable for all (perturbative) values of the right handed neutrino mass. 
For relatively light right handed neutrinos, their contribution to the renormalisation group flow is negligible.

\section{Reheating constraints and inflationary predictions\label{sec:inflation}}





\begin{table*}[t]
\begin{tabular}{c|ccccccc}
  $M_R$ (GeV)&$\xi$&$N_e$&$n_s$&$10^3\times r$&$10^4\times\alpha_s$&$10^2\times\phi_e/M_{\rm P}$&$10\times\phi_k/M_{\rm P}$\\
  \hline\\
  $2.3\times 10^{14}$ & ~$6501.4^{+689.5}_{-782.2}$ & $58.7759^{+0.0002}_{-0.0001}$ & $0.967172^{+0.000001}_{-0.000002}$ & $3.0833^{+0.0007}_{-0.0004}$ & $-5.4500^{+0.0004}_{-0.0006}$ & $1.3267^{+0.0873}_{-0.0648}$ & $1.1285^{+0.0751}_{-0.0556}$ \\\\
  $1000$ & $10951^{+379}_{-397}$ & ~$58.7788^{+0.0003}_{-0.0002}$ & ~$0.967148^{+0.000002}_{-0.000002}$ & ~$3.0895^{+0.0006}_{-0.0005}$ & ~$-5.4564^{+0.0005}_{-0.0006}$ & ~$1.0420^{+0.0210}_{-0.0189}$ & ~$0.88395^{+0.01769}_{-0.01592}$\\\\
  \hline
\end{tabular}
\caption{Prediction of the inflationary model for the two benchmark parameters $M_R=2.3\times 10^{14}$ GeV and $M_R=1$ TeV. 
}
\label{tab:Params}
\end{table*}

The Higgs quartic coupling of our model is seen to remain positive all the way up to the Planck scale and thus it is safe to use the Higgs effective potential for describing the dynamics of cosmic inflation. 
Including a nonminimal coupling between the Higgs field $H$ and the scalar curvature of the background geometry, the action pertinent to the inflationary dynamics reads, in the Jordan frame,
\begin{align}
  S=&\int d^4x\sqrt{-g}\Bigg\{
  \left(\frac{M_{\rm P}^2}{2}+\xi H^\dagger H\right)R\crcr
  &-(D_\mu H)^\dagger (D^\mu H)
  -V_{\rm eff}(H^\dagger H)\Bigg\},
\end{align}
where $M_{\rm P}=2.435\times 10^{18}$ GeV is the reduced Planck mass and 
\begin{align}\label{eqn:V}
  V_{\rm eff}(H^\dagger H)
  =\lambda\left(H^\dagger H-\frac{v^2}{2}\right)^2
\end{align}
is the Higgs potential.
The Higgs doublet in the unitary gauge is
\begin{align}
	H=\frac{1}{\sqrt 2}\begin{pmatrix}
		v+\phi\\ 0
	\end{pmatrix},
\end{align}
and the (physical) Higgs scalar $\phi$ is the inflaton of our model.
Identifying the scale of the running quartic coupling $\lambda$ with the inflaton value $\phi$, the renormalisation group improved effective potential for the inflaton becomes
\begin{align}\label{eqn:Veff}
  V_{\rm eff}(\phi)= \frac{\lambda(\phi)}{4}\phi^4,
\end{align}
where $v$ ($=246$ GeV) has been ignored as it is much smaller than the scale of inflation.

The dynamics of inflation are conveniently analysed in the Einstein frame by performing the Weyl transformation of the metric $g_{\mu\nu}\to g_{\mu\nu}^{\rm E}=\Omega(\phi)\, g_{\mu\nu}$, where
\begin{align}
  \Omega(\phi)=\sqrt{1+\xi\frac{\phi^2}{M_{\rm P}^2}}.
\end{align}
The canonically normalised inflaton field in the Einstein frame is denoted by $\sigma$ and is related to $\phi$ via
\begin{align}
  d\sigma = \frac{d\phi}{\Omega(\phi)^2}\sqrt{1+(1+6\xi)\xi\frac{\phi^2}{M_{\rm P}^2}}.
\end{align}
As discussed in \cite{George:2013iia,Kawai:2023dac}, it is natural to consider that dimensionful quantities are appropriately rescaled under the Weyl transformation, more specifically,
\begin{align}\label{eqn:Phi}
	\phi\to\Phi\equiv\frac{\phi}{\Omega(\phi)},
\end{align}
and the inflaton action and potential in the Einstein frame are respectively
\begin{align}
	S_{\rm E} =& \int d^4x\sqrt{-g_{\rm E}}\left\{
	\frac{M_{\rm P}^2}{2}R_{\rm E}-\half (\partial\sigma)^2-V_{\rm E}(\phi(\sigma))\right\},
\end{align}
\begin{align}\label{eqn:VE}
  V_{\rm E}(\phi)=\frac{\lambda(\Phi)}{4}\Phi^4.
\end{align}
Here $g_{\rm E}$ is the determinant of the Einstein frame metric $g_{\mu\nu}^{\rm E}$ and $R_{\rm E}$ is the corresponding scalar curvature.

The slow roll parameters are expressed using the original (Jordan frame) field $\phi$ as
\begin{align}
  \epsilon_V =& \frac{M_{\rm P}^2}{2}\left(
  \frac{V_{{\rm E},\sigma}}{V_{\rm E}}\right)^2
  =\half\left(\frac{M_{\rm P}}{\sigma_{,\phi}}\right)^2\left(
  \frac{V_{{\rm E},\phi}}{V_{\rm E}}\right)^2,\label{eqn:epsilonV}\\
  \eta_V =& M_{\rm P}^2
  \frac{V_{{\rm E},\sigma\sigma}}{V_{\rm E}}
  =\left(\frac{M_{\rm P}}{\sigma_{,\phi}}\right)^2\left(
  \frac{V_{{\rm E},\phi\phi}}{V_{\rm E}}
  -\frac{\sigma_{,\phi\phi}}{\sigma_{,\phi}}\frac{V_{{\rm E},\phi}}{V_{\rm E}}\right),\label{eqn:etaV}\\
  \zeta_V =& M_{\rm P}^4
  \frac{V_{{\rm E},\sigma}V_{{\rm E},\sigma\sigma\sigma}}{V_{\rm E}^2}\crcr
  &=\left(\frac{M_{\rm P}}{\sigma_{,\phi}}\right)^4
  \frac{V_{{\rm E},\phi}}{V_{\rm E}}
  \Bigg\{
  \frac{V_{{\rm E},\phi\phi\phi}}{V_{\rm E}}
  -3\frac{\sigma_{,\phi\phi}}{\sigma_{,\phi}}\frac{V_{{\rm E},\phi\phi}}{V_{\rm E}}\crcr
  &\qquad+\left(3\left(\frac{\sigma_{,\phi\phi}}{\sigma_{,\phi}}\right)^2-\frac{\sigma_{,\phi\phi\phi}}{\sigma_{,\phi}}\right)\frac{V_{{\rm E},\phi}}{V_{\rm E}}\Bigg\}.\label{eqn:zetaV}
\end{align}
By numerically solving the equation of motion for the inflaton field $\phi$ we evaluate the slow roll parameters, identify the end of inflation by the condition $\epsilon_V=1$, and find the prediction of the inflationary model.

If we know the value of the inflaton field $\phi=\phi_k$ at the time the comoving CMB scale $k$ exits the horizon, the corresponding amplitude of the curvature perturbation may be evaluated under the slow roll approximation as
\begin{align}\label{eqn:As}
  P_R = \left.\frac{V_{\rm E}}{24\pi^2M_{\rm P}^4\epsilon_V}\right\vert_{\phi=\phi_k}.
\end{align}
Comparing this with the measurements of the CMB (we use the
Planck 2018 TT, TE, EE + lowE + lensing central value \cite{Planck:2018jri} 
$\ln(10^{10}A_s) = 3.044$ at $k=0.05 \,\text{Mpc}^{-1}$ in actual computations), we fix the parameter $\xi$ of the nonminimal coupling.
Once this is done, the scalar spectral index, the tensor-to-scalar ratio and the running of the scalar spectral index at the pivot scale are evaluated using the slow roll parameters at $\phi=\phi_k$,
\begin{align}
  &n_s = 1-6\epsilon_V+2\eta_V,\quad
  r = 16\epsilon_V,\crcr
  &\alpha \equiv\frac{dn_s}{d\ln k} = -24\epsilon_V^2+16\epsilon_V\eta_V-2\zeta_V.
\end{align}

Traditionally, the inflaton value $\phi_k$ at the horizon exit of the CMB scale is fixed by solving the equation
\begin{align}\label{eqn:NkI}
  N_k=\frac{1}{M_{\rm P}}\int_{\phi_{\rm e}}^{\phi_k}
  \frac{d\phi}{\sqrt{2\epsilon_V(\phi)}}
  \left(\frac{d\sigma}{d\phi}\right)
\end{align}
for a given e-folding number $N_k$ between the horizon exit and the end of inflation (when $\phi=\phi_{\rm e}$).  
As emphasised recently in \cite{Kawai:2021hvs,Kawai:2023dac}, the e-folding number $N_k$ is not a free parameter but must be determined by the particle physics of the model\footnote{
See \cite{Bezrukov:2008ut,Garcia-Bellido:2008ycs} for studies of the original Higgs inflation scenario.
}.
This point has been overlooked, at least for this particular type of Higgs inflation model analysed in our previous work \cite{Okada:2015zfa}, and below we tie up the loose ends.

Let us first consider the perturbative decay scenario of reheating.
After inflation, the inflaton field starts to oscillate in the potential \eqref{eqn:VE}, which is essentially quartic $\propto\phi^4$ for small amplitudes $|\phi|\lesssim M_{\rm P}/\sqrt\xi$, see Eq. \eqref{eqn:Phi}.
In this regime the Universe expands as if radiation dominated, $a\propto \sqrt t$.
As the oscillation amplitude of the inflaton diminishes, the symmetry breaking effect, that is, the $v$ term in the potential \eqref{eqn:V}, becomes non-negligible and the equation of state changes, such that the Universe starts to expand as if matter dominated. 
At the moment of this transition, the energy density of the Universe may be evaluated as 
$\rho=\rho_\star\equiv\lambda(v) v^4/4 = m_h^2 v^2/8\sim 1.19\times 10^8\, \text{GeV}^4$.
The Hubble expansion rate at this moment is $H_\star=\sqrt{\rho_\star/3M_{\rm P}^2}=2.58\times 10^{-6}$ eV.
In our model the inflaton is the SM Higgs and its total decay width is known to be $\Gamma_h = 4.07$ MeV \cite{ParticleDataGroup:2022pth}. 
This is obviously much larger than $H_\star$, indicating that the inflaton already decays during oscillations in the quartic potential, and thermalisation would complete as soon as the expansion of the Universe resembles matter domination.
Beyond perturbative considerations, it has been pointed out \cite{Ema:2016dny} that preheating in this type of inflationary model can be very violent due to spike-like instabilities caused by the nonminimal coupling, and thus the inflaton may decay entirely after only a few oscillations at the end of inflation.
In either picture, reheating is completed when the Universe is in a radiation-domination-like expansion, and the evolution continues smoothly to the radiation dominant Universe.

This observation constrains the e-folding number as follows.
The ratio of the wave number $k/a_0$ for the present CMB scale to the Hubble parameter today $ H_0 = 100\,h\, {\rm km}\,{\rm s}^{-1}\,{\rm Mpc}^{-1}$, where $h=0.674$ \cite{Planck:2018vyg}, can be written as
\begin{align}
  \frac{k}{a_0H_0}=\frac{a_kH_k}{a_0H_0}
  =\frac{a_k}{a_{\rm e}}\frac{a_{\rm e}}{a_{\rm th}}
  \frac{a_{\rm th}}{a_{\rm eq}}\frac{a_{\rm eq}}{a_0}\frac{H_k}{H_0}.
\end{align}
Here, $H_k$ is the Hubble parameter at the horizon exit of the comoving scale $k$, and $a_k$, $a_{\rm e}$, $a_{\rm th}$, $a_{\rm eq}$, $a_0$ denote the scale factor at the horizon exit, end of inflation, thermalisation (completion of reheating), radiation-matter equality and the present Universe, respectively.
The e-folding number of inflation is minus the natural logarithm of the first factor, namely 
\begin{align}\label{eqn:Nk}
  N_k &\equiv \ln\frac{a_{\rm e}}{a_k}=66.5-\ln h-\ln\frac{k}{a_0H_0}
+\frac{1-3w}{12(1+w)}\ln\frac{\rho_{\rm th}}{\rho_{\rm e}}\crcr
&+\frac 14\ln\frac{V_k}{\rho_{\rm e}}
+\frac 14\ln\frac{V_k}{M_{\rm P}^4}
+\frac{1}{12}\left(\ln g_*^{\rm eq}-\ln g_*^{\rm th}\right),
\end{align}
where $w$ is the equation of state parameter during reheating, 
$\rho_{\rm th}$ and $\rho_{\rm e}$ denote the energy density of the Universe at thermalisation and at the end of inflation, $V_k$ is the value of the potential \eqref{eqn:VE} at the horizon exit $\phi=\phi_k$, and $g_*^{\rm eq}$ and $g_*^{\rm th}$ are the number of relativistic degrees of freedom at radiation-matter equality and thermalisation, respectively.  
The feature of this model mentioned above, that reheating is completed during the radiation-dominated-like expansion $w=1/3$, points to the fact that the term $\ln(\rho_{\rm th}/\rho_{\rm e})$ in \eqref{eqn:Nk} vanishes, and the e-folding number is determined entirely by the classical inflaton dynamics through $V_k=V_{\rm E}(\phi_k)$ and $\rho_{\rm e}\simeq 2V_{\rm E}(\phi_{\rm e})$. 
In particular, the reheating temperature $T_{\rm R}$ is not directly related to this parameter since $\rho_{\rm th}=\pi^2 g_*^{\rm th} T_{\rm R}^4/30$ has disappeared from \eqref{eqn:Nk}.

\begin{figure}
\includegraphics[width=85mm]{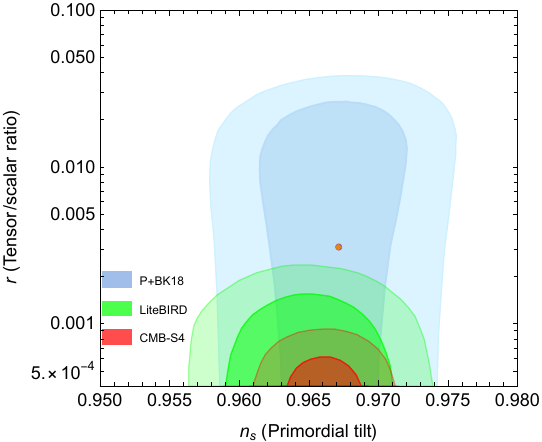}%
\caption
{\label{fig:nsrprediction}
The scalar spectral index $n_s$ and the tensor-to-scalar ratio $r$ predicted by the Higgs inflation model with minimal fermionic dark matter and three right handed neutrinos.
The purple bullet is for the right handed neutrino mass $M_{R}=2.3\times 10^{14}$ GeV, and the orange diamond is for $M_{R}=1$ TeV.
The light blue background contours are the 68\% and 95\% confidence level
Planck+BICEP/Keck 2018 constraints \cite{BICEP:2021xfz}.
The LiteBIRD \cite{LiteBIRD:2022cnt} (green) and CMB-S4 \cite{Abazajian:2019eic} (red) 1- and 2-$\sigma$ projective constraints for a fiducial model with $r=0$ are also shown.
}
\end{figure}

The requirement that the e-folding number $N_k$ determined by the field evolution \eqref{eqn:NkI} must coincide with \eqref{eqn:Nk} is a consistency condition that was missing in the previous analysis and we now implement this numerically.
To find the prediction of the inflationary model we solved the slow roll equation for the inflaton with the renormalisation group improved effective action.
With this consistency condition, there remains only one free parameter, the right handed neutrino mass $M_R$ in the model since the mass of the dark matter fermion has been fixed uniquely by the condition that it constitutes the entire dark matter abundance.
Table~\ref{tab:Params} lists the results for $M_R=2.3\times 10^{14}$ GeV and $1$ TeV, since these are roughly the upper and lower bounds from the perturbativity limit of the Dirac Yukawa coupling and from the collider experiments.
The uncertainties from the SM parameter measurements have negligible effects on these cosmological parameters, compared to the effect of choosing the different right handed neutrino mass parameter.
The predictions of the scalar spectral index $n_s$ and the tensor-to-scalar ratio $r$ are shown in Fig.~\ref{fig:nsrprediction}, with the background of the 68\% and 95\% confidence level contours from the Planck+BICEP/Keck 2018 data \cite{BICEP:2021xfz} (light blue), and projected constraints from LiteBIRD \cite{LiteBIRD:2022cnt} (green) and CMB-S4 \cite{Abazajian:2019eic} (red), for a $r=0$ fiducial model.
The prediction of the model is seen to be in an excellent fit with the Planck+BICEP/Keck data, and is clearly outside the 2-$\sigma$ contours of LiteBIRD and CMB-S4 with the $r=0$ fiducial model.
The two data points for $M_{R}=2.3\times 10^{14}$ GeV and $M_{R}=1$ TeV are nearly indistinguishable, indicating that the prediction of the model is quite robust.

In the presence of violent preheating \cite{Ema:2015dka}, the inflaton decays almost immediately after inflation and the reheating temperature is estimated as
\begin{align}
	T_{\rm R} \simeq \left(\frac{30\rho_e}{\pi^2 g_*^{\rm th}}\right)^\quarter\sim 3\times 10^{15}\;\text{GeV}.
\end{align}
The high reheating temperature supports thermal production of right handed neutrinos \cite{Davidson:2002qv} and the observed baryon asymmetry of the Universe can be produced via leptogenesis \cite{Fukugita:1986hr}.

\section{Final remarks\label{sec:final}}

As the Higgs boson of the SM is the only fundamental scalar field discovered so far in Nature, it is natural to speculate that it might play the role of the inflaton field required for the implementation of inflationary cosmology.
The current experimental values of the top quark mass and the Higgs boson mass however suggest that the Higgs potential becomes unstable for energy scale/field values $\gtrsim 10^{10}$ GeV, and thus becomes unsuitable to realise inflation.
An approach that is sometimes followed in order to circumvent this issue is to assume an appropriately small value for the top quark mass ($\gtrsim 3\sigma$), such that the Higgs potential becomes barely stable \cite{Bezrukov:2014bra,Hamada:2014wna}. 
Alternatively, one may accept the current central value of the top quark mass $m_t = 172.69$ GeV and assume, as we do, that the Higgs potential is stabilised due to the presence of a suitable dark matter particle.
Following this latter approach, we have analysed a cosmological scenario based on the so-called minimal dark matter model. 
In this paper we improved a previous analysis \cite{Okada:2015zfa} by including the consistency condition for reheating that has been overlooked, and reinvestigated the model using the updated constraints on dark matter fermions. 
We focused on an $SU(2)_L$ quintet fermionic dark matter since the triplet fermion dark matter model is strongly disfavoured by the indirect dark matter search experiments. 
The predictions for the scalar spectral index and the tensor-to-scalar ratio turn out to be very robust with $n_s = 0.9672$ and $r=0.0031$ respectively.
These values are in an excellent fit with the present CMB observation data, and are within the target range of near future experiments including the LiteBIRD and CMB-S4.
Our model is arguably the very minimal model of cosmic inflation so far as the WIMP dark matter scenario is postulated.
A key element is the $SU(2)_L$ quintet fermion, which, while currently is consistent with experiments \cite{Cuoco:2017iax}, may be excluded by an indirect dark matter search with antiproton flux measurements. 
Thus our model is falsifiable and can be tested by both CMB and cosmic rays.


\begin{acknowledgments}
%
This work was supported in part by 
the National Research Foundation of Korea Grant-in-Aid for Scientific Research Grant No. NRF-2022R1F1A1076172 (S.K.) and
by the United States Department of Energy Grant No. DE-SC0012447 (N.O.).
\end{acknowledgments}


\appendix

\section{Renormalization group equations\label{sec:RGEs}}

The renormalisation group improved effective potential of the Higgs field is obtained by solving the coupled equations listed below.
The two loop beta functions of the SM are given, for example, in \cite{Machacek:1983tz,Machacek:1983fi,Machacek:1984zw,Arason:1991ic,Luo:2002ey}.
We have added 
contributions from the dark matter fermion and the right handed neutrinos in our model.

The equations for the gauge couplings are
\begin{align}
  \frac{dg_i}{d\ln\mu}
  =\frac{g_i^3}{16\pi^2}b_i
  +\frac{g_i^3}{(16\pi^2)^2}\left(
  \sum_{j=1}^3 b_{ij}g_j^2-c_i y_t^2-\widetilde c_i S_\nu\right),
\end{align}
where $S_\nu\equiv y_D^\dagger y_D$ and we use the $SU(5)$ normalisation $g_1=\sqrt{5/3}\, g_Y$.
The beta functions are
\begin{align}
  b_i =& \left(\frac{41}{10},-\frac{19}{6}+\Delta b_2, -7\right),\\
  b_{ij} =& \left(
  \begin{array}{ccc}
  \frac{199}{50} & \frac{27}{10} & \frac{44}{5}\\
  \frac{9}{10} & \frac{35}{6}+\Delta b_{22} & 12\\
  \frac{11}{10} & \frac{9}{2} & -26
  \end{array}\right),\\
  c_i =& \left(\frac{17}{10}, \frac{3}{2}, 2\right),
\end{align}
where $\Delta b_2$ and $\Delta b_{22}$ are the contributions from the dark matter fermion.
Namely,
$\Delta b_2 = \frac{20}{3}$, $\Delta b_{22}=\frac{560}{3}$ for $\mu > M_{\rm DM}$, and
$\Delta b_2 = \Delta b_{22} = 0$ otherwise. 
The contribution $\widetilde c_i$ is from the right handed neutrinos,
\begin{align}
  \widetilde c_i =& \left(\frac{3}{10}, \half, 0\right),
\end{align}
for $\mu > M_R$, and $\widetilde c_i=0$ otherwise.

The flow of the top Yukawa coupling is given by
\begin{align}
  \frac{dy_t}{d\ln\mu} = y_t\left(
  \frac{\beta_t^{(1)}}{16\pi^2}+\frac{\beta_t^{(2)}}{(16\pi^2)^2}\right),
\end{align}
where the beta functions are
\begin{align}\label{eqn:betat1}
  \beta_t^{(1)} =& \frac 92 y_t^2 - \frac{17}{20}g_1^2 
  - \frac 94 g_2^2 - 8g_3^2 + S_\nu,\\
  \beta_t^{(2)} =& -12y_t^4 + \left(
  \frac{393}{80} g_1^2+\frac{255}{16} g_2^2+36*g_3^2\right)y_t^2\crcr
  &+\frac{1187}{600} g_1^4 - \frac{9}{20}g_1^2 g_2^2 
  +\frac{19}{15} g_1^2 g_3^2 -\frac{23}{4} g_2^4 \crcr
  &+9g_2^2g_3^2 -108g_3^4 +6\lambda^2-12\lambda y_t^2.
\end{align}
Here we only considered the contributions from the top quark .
The term involving $S_\nu$ in \eqref{eqn:betat1} (also in \eqref{eqn:betalambda1} below) is the contribution from the right handed neutrinos which vanishes for $\mu < M_R$.

The equation for the Higgs quartic coupling is
\begin{align}
  \frac{d\lambda}{d\ln\mu} = \frac{\beta_\lambda^{(1)}}{16\pi^2}
  +\frac{\beta_\lambda^{(2)}}{(16\pi^2)^2},
\end{align}
where
\begin{align}\label{eqn:betalambda1}
  \beta_\lambda^{(1)} =& 
  24\lambda^2 - \left(\frac 95 g_1^2+9g_2^2\right)\lambda 
  + \frac{27}{200} g_1^4+\frac{9}{20} g_1^2g_2^2\crcr
  &+\frac 98 g_2^4+12 y_t^2\lambda-6y_t^4 + 4S_\nu\lambda -2S_\nu^2,
\end{align}
\begin{align}
  \beta_\lambda^{(2)} =& -312\lambda^3 
  +108\left(\frac{g_1^2}{5}+g_2^2\right)\lambda^2 \crcr
  &-\left(\frac{73}{8} g_2^4-\frac{117}{20} g_1^2g_2^2
  -\frac{1887}{200} g_1^4\right)\lambda - 3\lambda y_t^4 \crcr
  &+\frac{305}{16} g_2^6 -\frac{289}{80} g_1^2 g_2^4 
  -\frac{1677}{400} g_1^4 g_2^2 - \frac{3411}{2000} g_1^6 \crcr
  &-32 g_3^2 y_t^4 - \frac 85 g_1^2 y_t^4 - \frac 94 g_2^4 y_t^2 \crcr
  &+\left(\frac{17}{2}g_1^2+\frac{45}{2} g_2^2+80 g_3^2\right)\lambda y_t^2\crcr 
  &-\frac{3}{10} g_1^2\left(\frac{57}{10} g_1^2-21 g_2^2\right) y_t^2-144\lambda^2 y_t^2+30 y_t^6.\crcr
\end{align}

Finally, for $\mu > M_R$, the evolution of the Dirac Yukawa coupling is governed by
\begin{align}
	\frac{dS_\nu}{d\ln\mu}=\frac{S_\nu}{16\pi^2}\left\{
	6y_t^2 + 5S_\nu -\left(\frac{9}{10}g_1^2+\frac 92 g_2^2\right)\right\}.
\end{align}

To solve these equations we employed the boundary conditions at the top quark pole mass \cite{Buttazzo:2013uya}, 
\begin{align}
g_1(m_t) =& \sqrt\frac 53 \Bigg\{0.35830+0.00011 (m_t-173.34)\cr
&-0.00020 \left(\frac{m_W-80.384}{0.014}\right)\Bigg\},\\ 
g_2(m_t) =& 0.64779+0.00004 (m_t-173.34)\cr
&+0.00011 \left(\frac{m_W-80.384}{0.014}\right),\\ 
g_3(m_t) =& 1.1666+0.00314 \left(\frac{\alpha_s-0.1184}{0.0007}\right)\cr
&-0.00046 \left(m_t-173.34\right),\\ 
y_t(m_t) =& 0.93690+0.00556 (m_t-173.34)\cr
&-0.00042 \left(\frac{\alpha_s-0.1184}{0.0007}\right),\\ 
\lambda(m_t) =& 0.12604+0.00206 (m_h-125.15)\cr
&-0.00004 (m_t-173.34), 
\end{align}
where we used \cite{ParticleDataGroup:2022pth,ATLAS:2023lhg} 
$m_t = 172.69\pm 0.30$ GeV, 
$m_W = 80.377\pm 0.012$ GeV, 
$m_h = 125.25\pm 0.17$ GeV and
$\alpha_s \equiv \alpha_s(m_Z)= 0.1183\pm 0.0009$,
$m_Z=91.1876\pm 0.0021$ GeV.

%


\end{document}